# Structure, optical and magnetic properties of Fe doped, Fe+Cr co-doped ZnO nanoparticles

H. S. Lokesha[1*], P. Mohanty[1], C. J. Sheppard[1] and A. R. E. Prinsloo[1]

[1]*Cr Research Group, Department of Physics, University of Johannesburg, PO Box 524, Auckland Park, Johannesburg, South Africa*

[*]Corresponding author email: lokeshahs@uj.ac.za

**Abstract:**

The study particularly focuses on the effect of the Cr co-doping on the structural, optical and magnetic properties of $Zn_{0.99}Fe_{0.01}O$. $Zn_{0.99-x}Fe_{0.01}Cr_xO$ ($0 \leq x \leq 0.05$) nanoparticles were synthesized by the solution combustion method. Powder x-ray diffraction (XRD) analysis confirms all the samples have hexagonal wurtzite structures without any secondary phases present in the spectra. The average crystallite size ($D$) and microstrain ($\varepsilon$) of the samples were calculated using the Williamson–Hall relation and $D$ was found to be $13 \pm 1$ nm for $Zn_{0.99}Fe_{0.01}O$. In Cr co-doped samples, $D$ increase slightly with Cr content and are given by $9 \pm 1$, $10 \pm 1$, and $11 \pm 1$ nm for the samples with $x = 0.01$, $0.03$ and $0.05$, respectively; while the strain ($\varepsilon$) decreased with increase in Cr co-doping. The cell parameters obtained from Rietveld refinement, the lattice parameters ($a$ and $c$) and cell volume increases with additional Cr doping. Transmission electron microscopy (TEM) images of the samples indicate that particles are in the nano-regime and agglomerated. Particle sizes are found to be $14 \pm 1$, $11 \pm 1$ and $12 \pm 1$ nm for $Zn_{0.99-x}Fe_{0.01}Cr_xO$ with $x = 0$, $0.03$ and $0.05$, respectively. The substitution of $Fe^{3+}$ and $Cr^{3+}$ at $Zn^{2+}$ sites has an impact on the optical properties. The band-gap decreases from $3.296 \pm 0.002$ eV to $3.258 \pm 0.002$ eV with increase of Cr concentration. Photoluminescence (PL) of $Zn_{0.96}Fe_{0.01}Cr_{0.03}O$ revealed the presence of defects, the emission peaks at 410 nm and 513 nm are attributed to Zn vacancies ($V_{Zn}$) and singly ionized oxygen vacancies ($V_o^+$), respectively. The $M(\mu_o H)$ curves of $Zn_{0.99-x}Fe_{0.01}Cr_xO$ ($x$ = 0.03 and 0.05) measured at room temperature using a vibrating sample magnetometer (VSM) are found to be hysteretic, signifying room temperature ferromagnetism (RTFM). Maximum saturation magnetization, $0.67 \pm 0.01$ emu.g$^{-1}$, is observed in $Zn_{0.94}Fe_{0.01}Cr_{0.05}O$. The observed RTFM in Fe+Cr co-doped ZnO is explained by of bound magnetic polaron (BMP) mechanism, the BMPs are formed by $V_{Zn}$ and $V_o^+$ defects. This paper enhances the understanding of the origin of RTFM in Fe+Cr co-doped ZnO nanoparticles.

## 1. Introduction

Nanosized diluted magnetic semiconductors (DMS) have been attractive because of the significant magnetic and magneto-electric properties these materials demonstrate, including the localized *d* electrons of the magnetic ions couple with the electrons in the semiconducting band [1,2]. In addition, doping of transition metals (TM: Cr, Co, Fe, Ni and Mn) into ZnO results in a change in the band–gap energy, the tunability of the band–gap is desired for using the material in various optoelectronic devices [3,4]. Several reports have concluded the existence of room temperature ferromagnetism (RTFM) in ZnO systems doped with Cr, Fe, Ni, Mn Co and In [5–9]. First-principles calculations show that the magnetic configuration of these compounds is sensitive to the arrangement of the ions in ZnO, and the doping with TMs leads to ferromagnetism [10–12]. For Cr and Fe doped ZnO, it was found that the optical behavior and RTFM are sensitive to the synthesis method [7,8,13]. The introduction of the RTFM ordering in Fe doped ZnO has been attributed to the oxidation state of the Fe, thus whether it is $Fe^{2+}$, $Fe^{3+}$ or a combination of both, [14–16]. RTFM is enhanced with an increase in $Fe^{3+}$ dopant concentration [15]. In Cr doped ZnO, the observed RTFM is attributed to defects and oxygen vacancies created by the substitution of Cr ions into Zn sites and a the exchange interactions between Cr 3d and O 2p spin moments [17,18].

Moreover, TM ions co-doping of ZnO is considered as a promising method to achieve enhanced RTFM [19], as is reported for Fe-Cu [20], Cu-Cr [21], Co-Cu [22], Mn-Co [23], Fe-Ni [19], Cr-Co [24] and Cr-N [25] ions co-doped ZnO. The magnetic properties in TM co-doped ZnO are enhanced when compared with single dopant ZnO [19,21,24–27]. Other studies showed paramagnetic (PM) behavior in single TM doped ZnO, while the behaviour was changed to ferromagnetic (FM) at RT for TM co–doped ZnO [28,29]. It is also noted that Fe–Cr co-doped ZnO is interesting because the material exhibits a single phase and shows RTFM [30].

The present work focuses on the structure, optical and magnetic properties of $Zn_{0.99-x}Fe_{0.01}Cr_xO$ ($0 \leq x \leq 0.05$) synthesized using solution combustion method. The structure and morphology of the as-synthesized samples were analyzed by x-ray diffraction (XRD) and high resolution transmission electron microscopy (HR-TEM). The optical band-gap behaviour as function of Cr concentration is analyzed using diffuse reflectance spectroscopy (DRS) studies. Defects analysis was carried out by using photoluminescence. Room temperature ferromagnetism was investigated using a vibrating sample magnetometer (VSM). In the present study, the observed RTFM in the samples are not attributed to any secondary phases. For this reason, the nature of defects and the concentration thereof is thought to play a key role in initiating the RTFM observed in the Fe+Cr co-doped ZnO, The RTFM observed is explained in this study based on the bound magnetic polaron (BMP) mechanism.

## 2. Experiment details

### 2.1 Material synthesis

$Zn_{0.99-x}Fe_{0.01}Cr_xO$, with $x = 0$, 0.01, 0.03 and 0.05, samples were synthesized by the solution combustion process using nitrates of $Zn(NO_3)_2 \cdot 6H_2O$ (98%c), $NH_2CH_2COOH$ (99%), $Cr(NO_3)_3 \cdot 9H_2O$ (99%) procured from Sigma-Aldrich company and $Fe(NO_3)_3 \cdot 9H_2O$ (98% SDFCL company). Stoichiometric amount (oxidizer to fuel ratio is unity) of the nitrates dissolved in 30 ml double distilled water and was magnetically stirred to obtain a homogenous solution. Then the solution was placed into a muffle furnace pre-heated to $350 \pm 10$ °C. The solution becomes foamed and then ignited, final voluminous foamy product (ash) was obtained. The product was crushed into a fine powder using an agate mortar and pestle.

## 2.2 Characterization

The structure of the samples was verified by the x-ray diffraction (XRD) technique at room temperature. XRD patterns of the samples were recorded using a Phillips PAN analytical X-pert Pro X-ray diffractometer (Cu–$K\alpha$ with $\lambda = 1.54056$ Å) in the $2\theta$ range from 10 to 90°. Particle sizes of the samples were determined using a transmission electron microscope (TEM) micrographs (Model: JEM–2100). The C-coated Cu grids (sample holder) were used for TEM measurements. The samples chemical compositions were analyzed using energy dispersive x-ray spectroscopy (EDX), utilizing a detector of Oxford Instruments attached to the TEM. The diffuse reflectance spectra (DRS) were recorded using a laboratory spectrometer instrument in the wavelength range 250 to 750 nm [31]. The magnetic measurement was carried out at room temperature using a vibrating sample magnetometer (VSM) (Mode: Lake Shore–7410 series).

## 3 Results and discussion

### 3.1 X-ray diffraction

Rietveld refinement of XRD patterns of as-prepared $Zn_{0.99-x}Fe_{0.01}Cr_xO$, with $x = 0$, 0.01, 0.03 and 0.05, samples are shown in figure 1. XRD results indicate that the as-prepared samples are polycrystalline in nature. All the peaks are characteristic of the hexagonal wurtzite structure of ZnO with space group P63mc (PDF#36-1451). There are no secondary phases (such as $ZnFe_2O_4$, $Fe_2O_3$, $Fe_3O_4$ or $ZnCrO_4$ phases) observed in both Fe doped and Fe−Cr co-doped ZnO samples within the detection limit of XRD. The enlarged region of the (100), (002) and (101) diffraction peaks of $Zn_{0.99-x}Fe_{0.01}Cr_xO$ ($x = 0$, 0.01, 0.03 and 0.05) are shown in figure 2(a), indicates that the full width half maximum (FWHM) slightly decreases with the increase of Cr concentration with a negligible peak shift in $2\theta$ position. It also confirms that the doped Fe and Cr substitute into Zn sites without altering the ZnO phase. Most probably the $Fe^{3+}$ (0.62 Å) and $Cr^{3+}$ (0.61 Å) ions replace the $Zn^{2+}$ (0.74 Å) in the ZnO structure [32,33]. The average crystallite size (*D*) and microstrain (*ε*) were calculated using Williamson-Hall (W−H) method [34], the details of the method was reported elsewhere [35]. Figure 2(b) shows the plot of $(\beta \cos\theta/\lambda)$ as a function of $(4\sin\theta/\lambda)$, where *β* is the full width half maximum (FWHM) and *λ* is the wavelength of x-ray. The '*D*' and '*ε*' were calculated from the inverse of the *y*-intercept and slope of the linear fit of data, respectively. The calculated structural parameters are tabulated in Table 1. It is found that, the crystallite size is slightly enhanced from 9 ± 1 nm to 11 ± 1 nm and microstrain decreases with the increase of Cr concentration.

Further, the unit cell parameters were obtained from Rietveld refinement of XRD using the GSAS II software program [36]. The refined and fitted parameters are given in Table 1. The fitting parameter $\chi^2 \leq 1.2$, indicating that both the theoretical and experimental data are well matched with each other and the obtained lattice parameters are comparable with standard data (PDF#36-1451). However, it is noted that the lattice parameters *a* and *c*, as well as the cell volume are slightly enhanced in $Zn_{0.99-x}Fe_{0.01}Cr_xO$ (with *x* = 0.01, 0.03 and 0.05) as compared to the original $Zn_{0.99}Fe_{0.01}O$ sample. The expansion of the lattice parameters and cell volume is attributed to lattice distortions and defects created around dopants sites due to differences in ionic radii of the various cations (Zn, Fe and Cr) [37,38].

*3.2 Transmission electron microscope*

TEM images of $Zn_{0.99-x}Fe_{0.01}Cr_xO$ with *x* = 0, 0.03 and 0.05, are depicted in figure 3 (a) to (c). The corresponding HR-TEM image and particles distribution of the samples are shown in figure 3 (d) to (i). It is observed that for all the samples the particles are slightly agglomerated and nearly spherical or elliptical in shape but non-uniform size. The particle size was estimated by considering several TEM images and the size distribution bar graphs are well fitted with log-normal function. The average particle size is found be 14 ± 1, 11 ± 1 and 12 ± 1 nm for $Zn_{0.99-x}Fe_{0.01}Cr_xO$ with $x$ = 0, 0.03 and 0.05, respectively. From the HR-TEM images, the spacing between two adjacent fringes ($d$-spacing) was calculated and found to be 0.286 ± 0.003 nm for $Zn_{0.99}Fe_{0.01}O$ corresponding to the spacing between the (100) planes of the hexagonal wurtzite structure of ZnO. The measured $d$-spacing value is 0.521 ±0.002 and 0.522 ±0.002 nm for $Zn_{0.99-x}Fe_{0.01}Cr_xO$ with $x$ = 0.03 and 0.05, this is close value of lattice constant ($c$ = 5.2238 ± 0.0005) of ZnO grown along the $c$-axis direction [39]. Energy dispersive x-ray spectroscopy (EDX) results provides information about the chemical compositions in the samples. The EDX results reveal the presence of Zn, O, Fe elements in $Zn_{0.99}Fe_{0.01}O$ and Zn, O, Fe and Cr elements in $Zn_{0.96}Fe_{0.01}Cr_{0.03}O$, which showed the homogenous distribution of dopants in ZnO host. There were no additional elements detected, excluding the Cu and C seen from the sample grid.

3.3 *Optical properties*

The DRS spectra were used to determine the band-gap of the samples. Figure 4(a) shows the DRS spectra of $Zn_{0.99-x}Fe_{0.01}Cr_xO$, with $x$ = 0, 0.01, 0.03 and 0.05, samples. It is observed that the reflectance intensity decreases and the absorption edge slightly shifted towards a higher wavelength with the increase of Cr concentration. The small absorption humps are noticed in the wavelength range from 410 nm to 600 nm, corresponding to $d$–$d$ transitions between $^6A_{1g}$ ground states to $^4T$, $^4E$, and $^4A_1$ excited states of $Fe^{3+}$ ions replacing $Zn^{2+}$ sites [30,40]. However, if $Fe^{2+}$ ions coexist with $Fe^{3+}$ ions, this cannot be detected by the DRS measurements. Also the small hump located at 618 nm in Fe+Cr co-doped ZnO samples is assigned to ($^4A_{2g}(F) \rightarrow {^4T_{2g}(F)}$) transition of $Cr^{3+}$ ions in the ZnO lattice [30]. The Kubelka−Munk theory [41] was applied to determine the band-gap of the samples as function Cr concentration. The band-gap values of the samples were obtained by extrapolating the linear region of $[F(R)h\nu]^2$ as a function of photon energy ($h\nu$) plots [41], where $F(R)$ is the Kubelka–Munk function $F(R) = \frac{(1-R)^2}{2R}$ and $R$ is the reflectance. The plot of $[F(R)h\nu]^2$ against the $h\nu$ of $Zn_{0.99-x}Fe_{0.01}Cr_xO$ (with $x = 0$ and 0.05) is shown in figure 4(b). The calculated band-gap value of the samples are given in table 1, the band-gap decreases from $3.296 \pm 0.002$ eV to $3.258 \pm 0.002$ eV with increase in Cr concentration from $x = 0$ to 0.05. Decrease of band-gap values with the increase in Cr concentration is attributed to *sp–d* exchange interaction between the band electrons of ZnO and localized *d* electrons of the Fe and Cr ions substituting for Zn ions [42].

In order to confirm the presence of intrinsic defects, PL spectrum of $Zn_{0.96}Fe_{0.01}Cr_{0.03}O$ was recorded at room temperature with an excitation wavelength 248 nm. Figure 5 shows the de-convoluted PL spectrum of $Zn_{0.96}Fe_{0.01}Cr_{0.03}O$, fitted into five peaks using Gaussian fits centered at 381, 410, 458, 513 and 606 nm. The peak in the UV region is detected at 381 nm is attributed to near band edge emission of ZnO, which originates from the recombination of free excitons from a localized level near the condition band to the valence band [43]. The emission peak at 410 nm corresponds to zinc vacancies ($V_{Zn}$) [44]. The blue-green band (458 nm) is the radiative transition from zinc interstitial ($Zn_i$) to acceptor level of neutral $V_{Zn}$ near valence band [45]. The emission at 513 nm is observed due to the recombination of electrons and holes trapped in single ionized oxygen vacancies ($V_o^+$) [42,44–46]. The emission peak around 606 nm is related to oxygen interstitials ($O_i$) defects [47]. The intrinsic defects such as $V_{Zn}$, $V_o^+$ and $O_i$ are formed in the synthesized $Zn_{0.99-x}Fe_{0.01}Cr_xO$ (with $x = 0$, 0.01, 0.03 and 0.05), owing to stabilization of the structure [48].

### 3.4 *Magnetic properties*

The magnetization as function of magnetic field, $M(\mu_o H)$ curves of $Zn_{0.99-x}Fe_{0.01}Cr_xO$, with $x = 0.03$ and $0.05$, measured at room temperature are depicted in figure 6. The enlarged view of the behaviour at lower fields is shown in inset figure 6. The $M(\mu_o H)$ curves of both samples are hysteretic, the coercive field ($H_c$) of the samples is $147 \pm 1$ Oe, indicating the RTFM ordering. It is found that, the saturation magnetization ($M_s$) is increased from $0.556$ emu.g$^{-1}$ to $0.852$ emu.g$^{-1}$, with an increase of Cr co-dopant concentration. The origin of RTFM in $Zn_{0.99-x}Fe_{0.01}Cr_xO$, with $x = 0.03$ and $0.05$, is possibly intrinsic because the PL results confirm the defects present in the sample such as $V_{Zn}$, $V_o^+$, and $O_i$. XRD and EDX analysis also conclude there is no existence of impurities or secondary phases in the samples. Karmakar et al. [49] reported that the Fe present in both valence states ($Fe^{2+}$ and $Fe^{3+}$) in the ZnO and the $Fe^{3+}$ valence state leads to Zn vacancy formation and it mediates the exchange interactions between the iron ions. Aljawfi et al. [24] reported the RTFM in Cr–Co doped ZnO, the exchange interactions between the dopants with mediated oxygen vacancies explained on the basis of bound magnetic polaron (BMP) mechanism. Moreover, using the cation density as $3.94 \times 10^{22}$ cm$^{-3}$ in ZnO, BMP concentration for percolation threshold around $2 \times 10^{18}$ cm$^{-3}$ required for long-range ferromagnetism [50,51].

In order to obtain a better insight of the role of defects in ordering the RTFM in $Zn_{0.99-x}Fe_{0.01}Cr_xO$ (with $x = 0.03$ and $0.05$), the experimental $M(\mu_o H)$ curve of the samples were fitted to the BMP model using the equation [51,52]:

$$M = M_0 L(x) + \chi_m H,$$

where $L(x)$ is the Langevin function with $x = m_{eff}H/k_BT$ and $M_0 = nm_s$ is the total BMP magnetization, $n, m_s, m_{eff}, k_B, \chi_m$ are the number of BMPs, the effective spontaneous moment per BMP, the true spontaneous moment per BMP, the Boltzmann constant and the susceptibility of the matrix, respectively. The first term signifies the contribution of BMP and the second term represents the paramagnetic matrix contribution. The experimental $M(\mu_oH)$ curve of the samples is closely fitted with the BMP model as shown in figure 7. The fitted parameters $M_0$ and $\chi_m$ are found to be 0.44 and 0.65 emu.g$^{-1}$, 3.3×10$^{-6}$ and 4.9×10$^{-6}$ cgs for $Zn_{0.99-x}Fe_{0.01}Cr_xO$ (with $x =$ 0.03 and 0.05), respectively. At higher temperatures it was found that $m_s \cong m_{eff}$, with a magnitude in the order of 10$^{-18}$ emu for both samples. The number of BMPs is determined to be 0.98×10$^{18}$ and 2.12×10$^{18}$ cm$^{-3}$ for $Zn_{1-x}Cr_xO$ ($x =$ 0.03 and 0.05), respectively. Here, the number of BMPs formed in $Zn_{0.96}Fe_{0.01}Cr_{0.03}O$ is relatively smaller than the threshold value. Thus, the scenario where the direct interaction of BMPs results in the long range FM ordering simply cannot be applicable here [53]. Some of the reports [30,42,55] showed the indirect interaction of BMPs, i.e. the interaction between BMPs and metallic clusters [54] or grain boundaries, are responsible for ordering the RTFM. In the current work, the presence of the secondary phase is ruled out, therefore the interactions of BMPs associated with oxygen vacancies and /or defects (present at the surface) are responsible for the observed RTFM in $Zn_{0.96}Fe_{0.01}Cr_{0.03}O$. In the case of $Zn_{0.94}Fe_{0.01}Cr_{0.05}O$, the number of BMPs (2.12×10$^{18}$ cm$^{-3}$) is greater than the threshold value (2×10$^{18}$ cm$^{-3}$) [50,51]. Therefore, the overlapping of BMPs causes the alignment of dopant spins, resulting the long-range ferromagnetism observed at room temperature [53]. Finally, it is concluded that the co-doping of Cr ions contribute to the formation of more defects without altering the crystal structure of the ZnO system, resulting in the formation of more BMPs and enhancing the ferromagnetism.

4. **Conclusions**

In summary, $Zn_{0.99-x}Fe_{0.01}Cr_xO$ (with $0 \leq x \leq 0.05$) nanoparticles, with hexagonal wurtzite structure, were synthesized by the solution combustion method. The average crystallite size of the samples in the range from $9 \pm 1$ nm to $13 \pm 1$ nm and microstrain decreases with co-doping and increase of the Cr concentration. The average particle size is found to be $14 \pm 1$ nm for $Zn_{0.99}Fe_{0.01}O$ and is reduced to $12 \pm 1$ nm for $Zn_{0.96}Fe_{0.01}Cr_{0.03}O$. The band-gap decreases with the increase of Cr concentration. The defects present in $Zn_{0.96}Fe_{0.01}Cr_{0.03}O$ were analyzed by PL and found to be $V_{Zn}$, $V_o^+$ and $O_i$ defects. The observed RTFM in $Zn_{0.99-x}Fe_{0.01}Cr_xO$ (with $x = 0.03$ and $0.05$) samples is intrinsic and is not attributed to any impurity phase – as is explained on the basis of the bound magnetic polaron (BMP) mechanism. These findings suggested that $Fe^{3+}$ and $Cr^{3+}$ ions are successfully incorporated into $Zn^{2+}$ sites of the wurtzite ZnO. At higher Cr concentration, the sufficient number of BMPs formed is greater than the number of BMPs required to achieve percolation threshold, because the presence of $V_{Zn}$ and $V_o^+$ defects. Hence, the long-range ferromagnetic ordering observed from the overlapping of BMPs causes the alignment of their spins in $Zn_{0.94}Fe_{0.01}Cr_{0.05}O$ sample.

**Acknowledgement**

Financial aid from the south African National Research Foundation (Grant No's: 120856 and 88080) and the Faculty of Science, University of Johannesburg (UJ) URC and FRC is acknowledged. The use of the NEP Cryogenic Physical Properties Measurement System at UJ, obtained with the financial support from the SANRF (Grant No: 88080) and UJ, RSA, is acknowledged. The use of the Spectrum Analytical Facility at UJ is acknowledged.

**Table 1**. Structural parameters of $Zn_{0.99-x}Fe_{0.01}Cr_xO$ ($x = 0$, 0.01, 0.03 and 0.05) samples estimated from Rietveld refinement of XRD using GSAS II program. The fitted parameters chi-square ($\chi^2$), Weighted Profile ($R_{wp}$) and scale F factor ($R_F$) are given in Table 1.

| Refined parameters | $Zn_{0.99-x}Fe_{0.01}Cr_xO$ | | | |
| --- | --- | --- | --- | --- |
| | $x = 0$ | $x = 0.01$ | $x = 0.03$ | $x = 0.05$ |
| Crystallite size (nm) from W-H method | 13 ± 1 | 9 ± 1 | 10 ± 1 | 11 ± 1 |
| Microstrain (%) | 0.153 | 0.126 | 0.051 | 0.048 |
| Particle size from TEM (nm) | 14 ± 1 | -- | 11 ± 1 | 12 ± 1 |
| Lattice parameters (Å) (Error:±0.0005) | $a = b = 3.2537$ $c = 5.2124$ | 3.2579 5.2214 | 3.2601 5.2238 | 3.2583 5.2196 |
| Cell volume (Å³) | 47.791 ± 0.007 | 47.99 ± 0.01 | 48.08 ± 0.01 | 47.99 ± 0.01 |
| $\chi^2$ | 1.14 | 1.04 | 1.06 | 1.11 |
| $R_{wp}$ (%) | 4.01 | 3.81 | 3.92 | 4.11 |
| $R_F$ (%) | 2.11 | 4.04 | 4.70 | 4.73 |
| Band-gap (eV) | 3.296 ± 0.002 | 3.285±0.002 | 3.278±0.002 | 3.258±0.002 |
| Zn $(x, y, z)$ | (0.333, 0.666, −0.007) | (0.333, 0.666, −0.011) | (0.333, 0.666, -0.004) | (0.333, 0.666, -0.002) |
| O $(x, y, z)$ | (0.333, 0.666, 0.379) | (0.333, 0.666, 0.666) | (0.333, 0.666, 0.384) | (0.333, 0.666, 0.385) |
| Fe $(x, y, z)$ | (0.333, 0.666, 0.081) | (0.333, 0.660, 0.404) | (0.333, 0.666, 0.398) | (0.333, 0.666, 0.400) |
| Cr $(x, y, z)$ | -- | (0.333, 0.660, 0.341) | 0.333, 0.660, 0.341) | 0.333, 0.660, 0.341) |

**Figures**

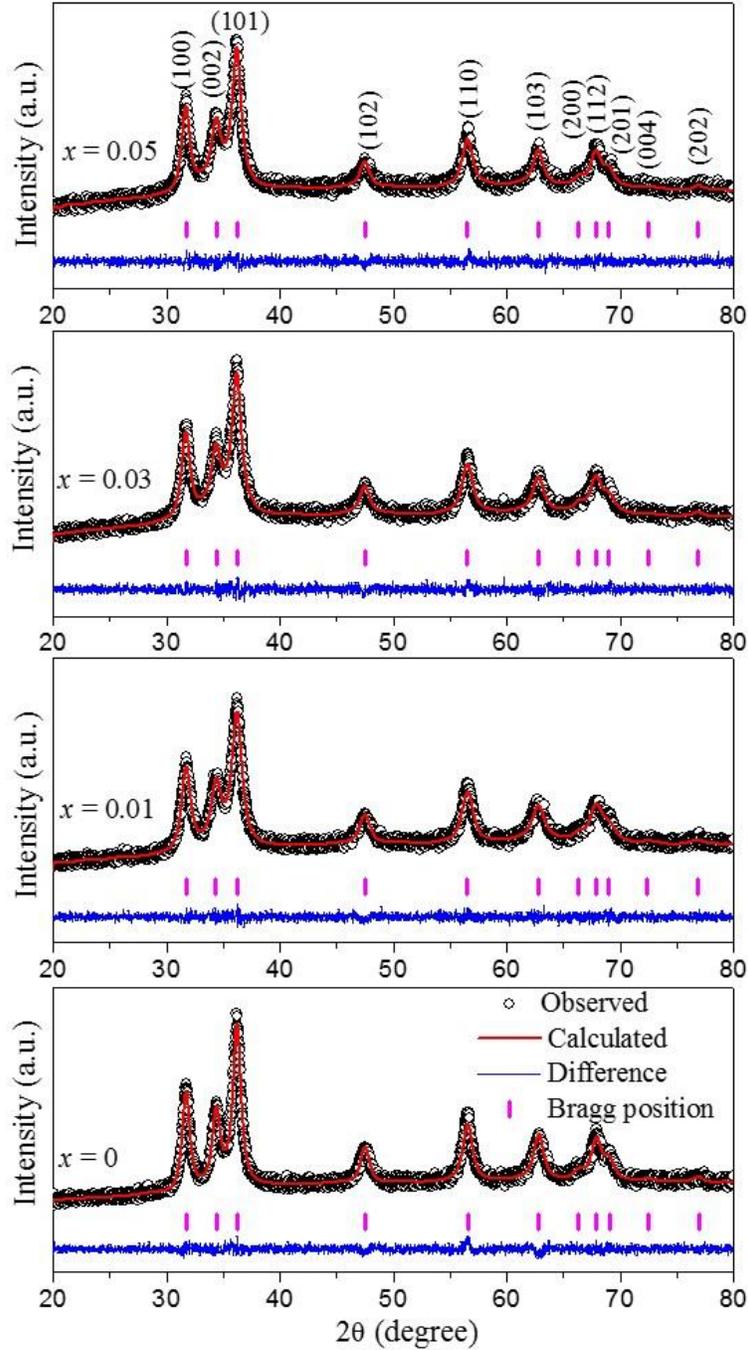

**Figure 1.** Rietveld refinement of XRD patterns of the $Zn_{0.99-x}Fe_{0.01}Cr_xO$ (with $x$ = 0, 0.01, 0.03 and 0.05) analyzed using GSAS II program. The black open circle represent the measured data compared with the calculated profile, the small purple vertical lines below the curve are the expected Bragg positions of wurtzite structure of ZnO and the residual of the refinement shown as blue solid line.

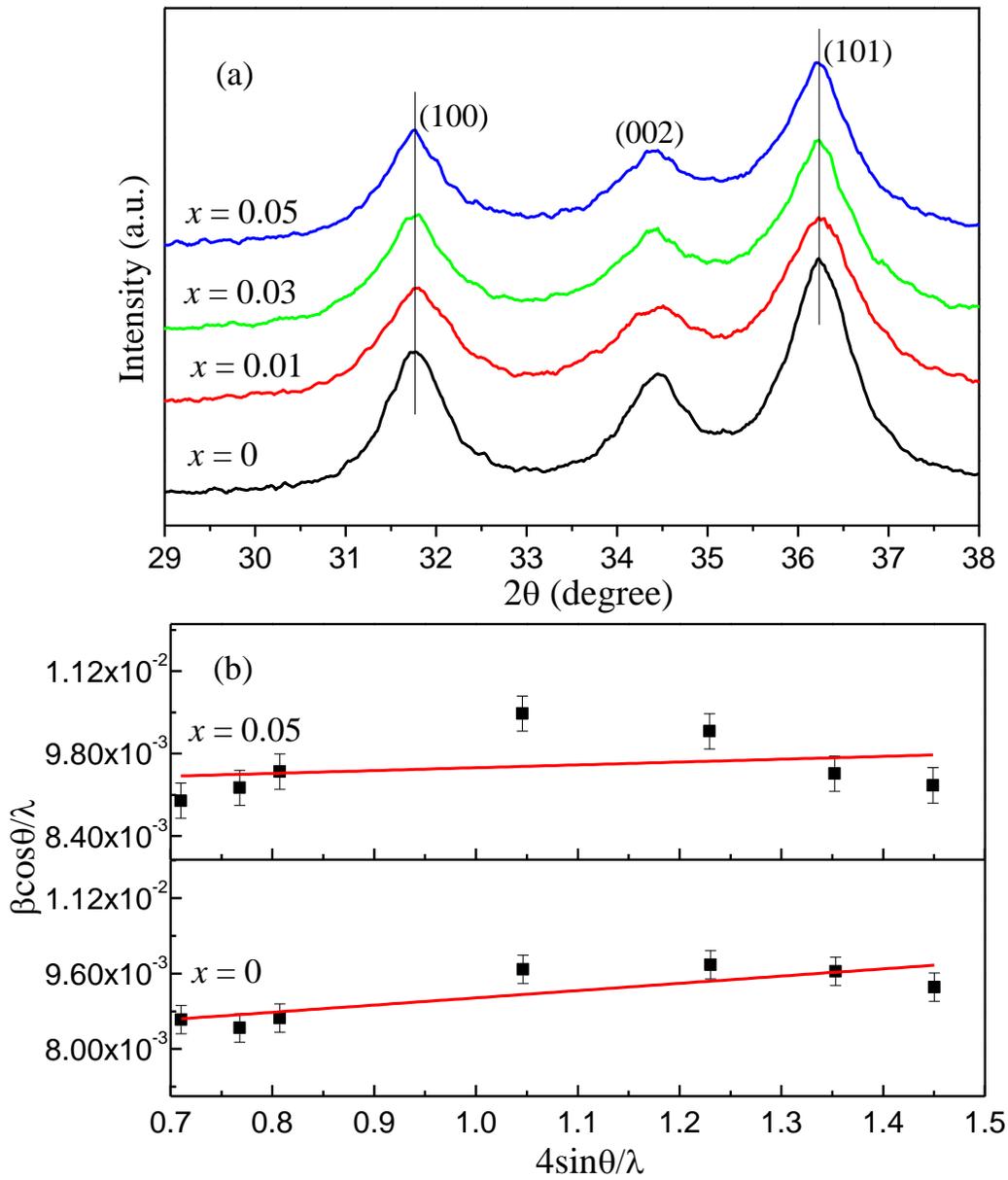

**Figure 2.** (a) The enlarged region of the (100), (002) and (101) diffraction peaks of $Zn_{0.99-x}Fe_{0.01}Cr_xO$ ($x$ = 0, 0.01, 0.03 and 0.05). (b) The Williamson−Hall plot, ($\beta\cos\theta/\lambda$) as function of ($4\sin\theta/\lambda$) for $Zn_{0.99-x}Fe_{0.01}Cr_xO$ ($x$ = 0 and 0.05) samples, where the red lines in this panel represents linear fit of the data.

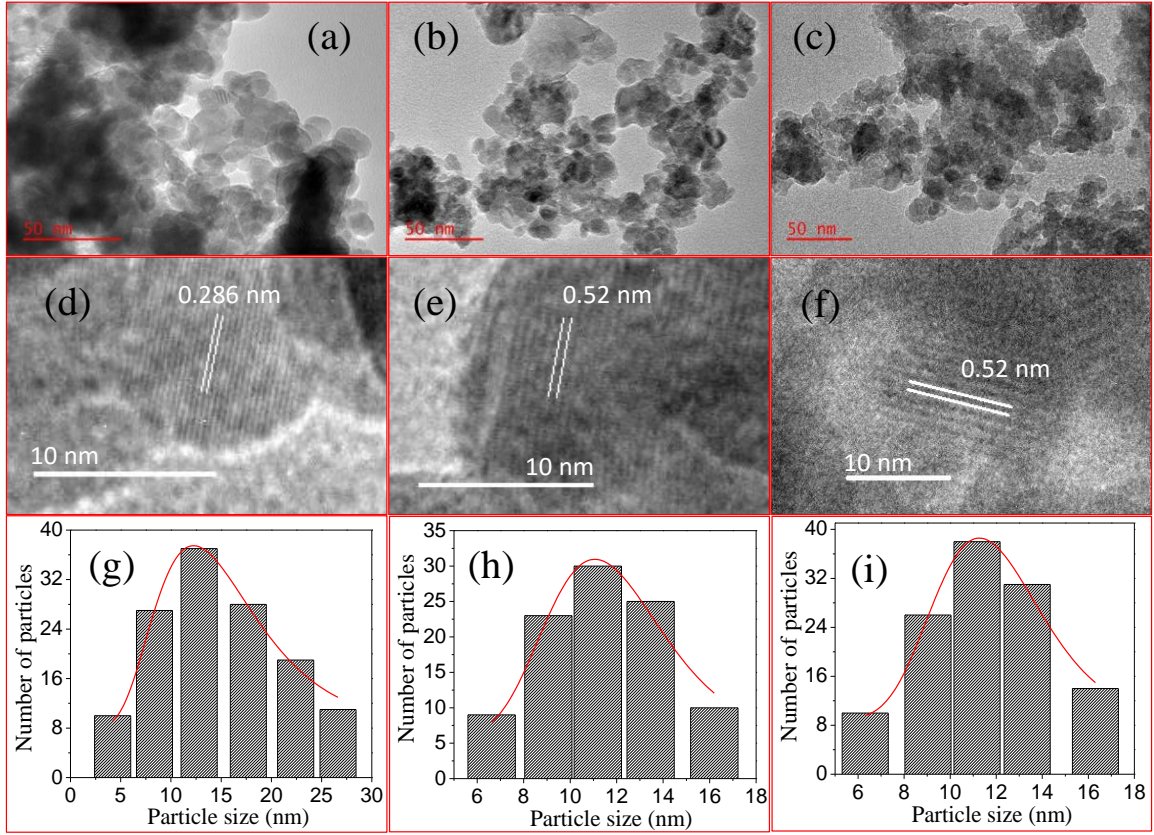

**Figure 3.** TEM images of (a) of $Zn_{0.99}Fe_{0.01}O$, (b) $Zn_{0.96}Fe_{0.01}Cr_{0.03}O$ and (c) $Zn_{0.94}Fe_{0.01}Cr_{0.05}O$ samples and their corresponding high resolution TEM images shown in figure 3 (d) to (f). The particle size distribution of the samples are given in figure 3 (g) to (i), respectively, where the red lines are log-normal fits to the data.

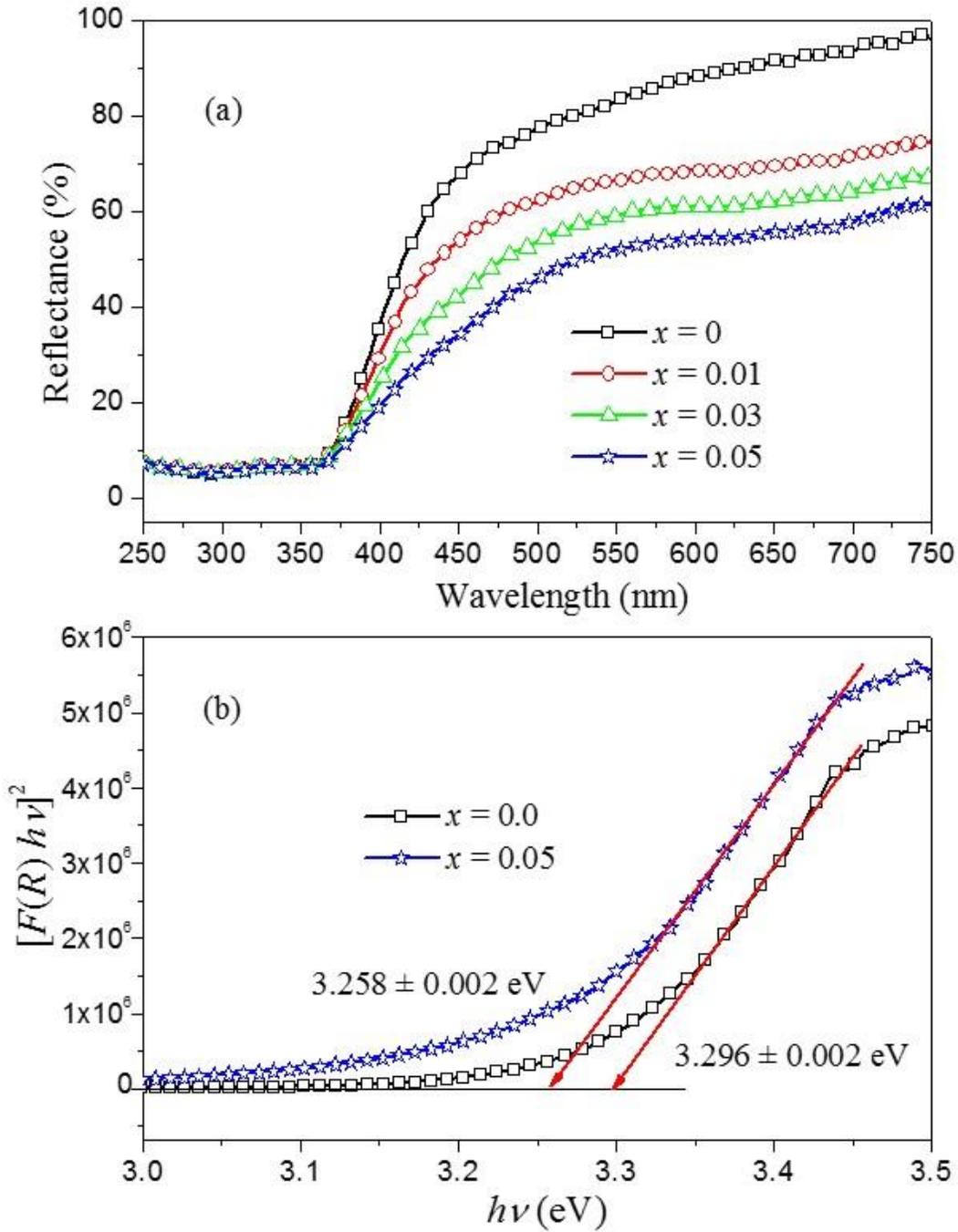

**Figure 4.** (a) DRS spectra of $Zn_{0.99-x}Fe_{0.01}Cr_xO$ ($x = 0$, 0.01, 0.03 and 0.05) samples. (b) The plot of $[F(R)h\nu]^2$ as function of $(h\nu)$ for $Zn_{0.99-x}Fe_{0.01}Cr_xO$ ($x = 0$ and 0.05) samples, the red lines represent the extended lines from the linear region.

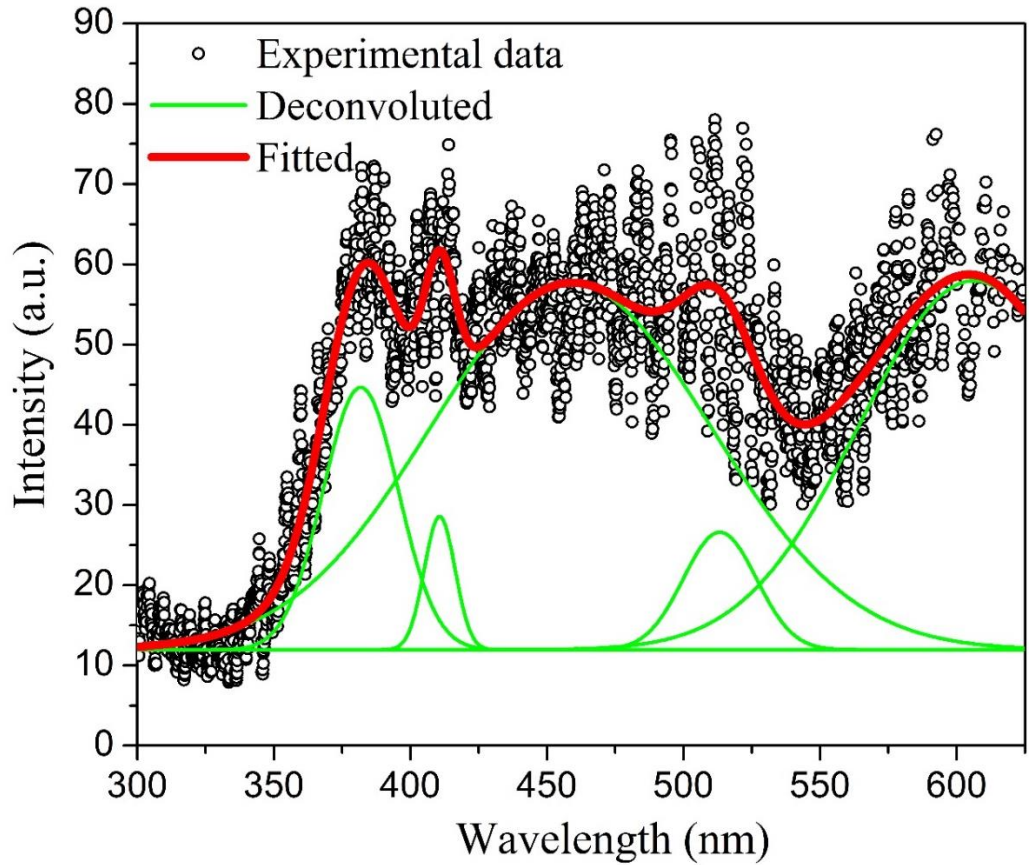

**Figure 5.** Photoluminescence emission spectrum of $Zn_{0.96}Fe_{0.01}Cr_{0.03}O$ measured at room temperature under excitation wavelength 248 nm. The spectrum de-convoluted into five peaks using Gaussian fits, the symbol is experimental data, the thick (red) solid line fitted and the thin (green) solid line is de-convoluted peaks.

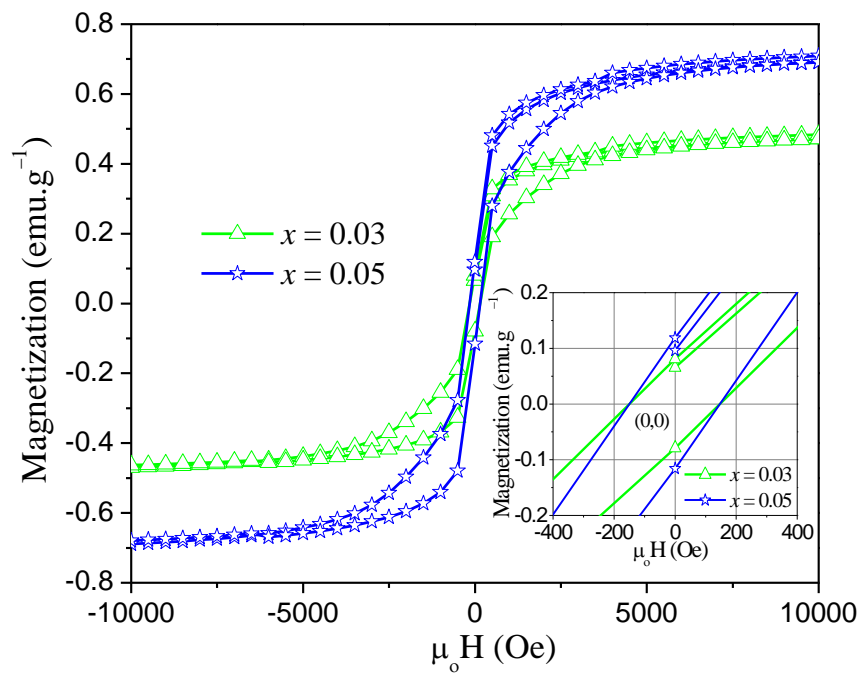

**Figure 6.** Magnetization as function of magnetic field of $Zn_{0.99-x}Fe_{0.01}Cr_xO$ ($x = 0.03$ and $0.05$) samples measured at room temperature. The inset is the enlarged $M(\mu_oH)$ curves at lower fields.

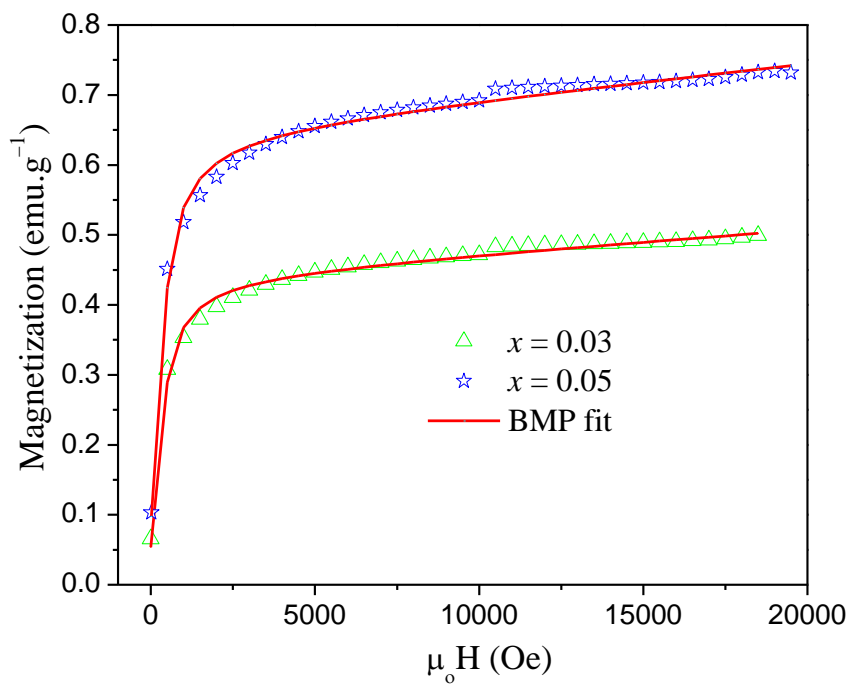

**Figure 7.** The symbols are experimental data of $M(\mu_oH)$ curves of $Zn_{0.99-x}Fe_{0.01}Cr_xO$ ($x = 0.03$ and $0.05$) and red line the fits to the experimental data using the BMP model.